\documentclass[prb,aps,twocolumn,showpacs,epsf]{revtex4}
\begin{document}

\title{Confinement of matter fields in compact (2+1)-dimensional QED theory of high-$T_{c}$ superconductors}
\author{Guo-Zhu Liu \\
{\small {\it Department of Mathematics, University of Science and
Technology of China, Hefei, Anhui, 230026, P.R. China }}}

\begin{abstract}
The confinement of matter fields is studied in the compact
QED$_{3}$ theory of high-$T_{c}$ superconductors. It is found that
the monopole configurations do not affect the propagator of gauge
potential $a_{\mu}$. This then leads to the findings that chiral
symmetry breaking and confinement take place simultaneously in the
antiferromagnetic state and that neither monopole effect nor
Anderson-Higgs mechanism can cause confinement in the $d$-wave
superconducting state. The physical implications of these field
theoretic results are also discussed.
\end{abstract}


\maketitle


The idea that high-$T_{c}$ superconductor is some kind of quantum
spin liquid \cite{Anderson} motivated much research effort in the
past seventeen years. Based on slave-boson treatment of $t$-$J$
model, it was found that the low-energy physics of the
antiferromagnetic Mott insulator is captured by a theory of
gapless Dirac fermions interacting with a U(1) gauge field
\cite{Ioffe, Marston89}. Doping drives the Mott insulator to a
$d$-wave superconductor, which can be described by a more general
U(1) gauge theory including both Dirac fermions and scalar bosons
\cite{Kim99}. The U(1) gauge field is not the usual
electromagnetic field \cite{Ioffe, Marston89, Kim99, Baskaran}. It
originates from the strong correlation effect and is obtained in
general by spontaneously breaking a larger SU(2) gauge symmetry.
It has been proved \cite{Affleck} that cuprate superconductors at
zero doping contain a local SU(2) gauge symmetry once
slave-particle approach and the constraint of one particle per
site are adopted. Later, Wen and Lee \cite{Wen96} constructed a
SU(2) gauge structure away from half-filling. In the staggered
flux phase, two components of the SU(2) gauge field become massive
via Higgs mechanism and hence are neglected \cite{Wen96}, leaving
a massless U(1) gauge field. As claimed by Polyakov
\cite{Polyakov77}, this U(1) gauge field must be $compact$ because
the SU(2) group is defined on a compact sphere. The compact gauge
structure also appears in theories of the Neel state and various
spin liquid phases of the planar quantum Heisenberg
antiferromagnets \cite{Sachdev, Senthil}. In particular, recently
it has been used to build a critical theory of zero-temperature
quantum phase transitions \cite{Senthil} that can not be described
by the conventional Wilson-Ginzburg-Landau paradigm.

Polyakov \cite{Polyakov77} found that the (2+1)-dimensional
compact quantum electrodynamics (QED$_{3}$) has monopole
configurations around which gauge potential $a_{\mu}$ jumps
$2n\pi$. The most remarkable effect of monopoles is that it leads
to permanent confinement of static charges \cite{Wilson}. In order
to understand realistic condensed matter systems, it is necessary
to couple $a_{\mu}$ to fermions and scalar bosons. However, though
the confinement in pure compact QED$_{3}$ is now widely accepted,
the confinement of dynamical matter fields is far from clear. The
confinement of massless fermions is of particular interests since
they exist in the whole phase diagram of cuprate superconductors
due to the $d$-wave gap symmetry. In our opinion, all previous
efforts \cite{Ioffe, Marston, Herbut03, Kleinert} towards the
confinement problem of massless fermions are not satisfactory
since they did not give a careful analysis of the relationship
between chiral symmetry breaking (CSB), monopoles and confinement.
Such an analysis is necessary not only in studying correlated
electron systems but in understanding QCD, the theory of strong
interactions.

In this paper we study the confinement of matter fields in the
compact QED$_{3}$ theory of the high-$T_{c}$ cuprate
superconductors. We concentrate on the half-filled
antiferromagnetic Mott insulator state and the $d$-wave
superconducting state, two most interesting ground states in
cuprate superconductors. Actually, the various strange behavior
that can not be understood within conventional many-body theory
are generally believed to arise from the competition between these
two orders. We make a consistent treatment of confinement with CSB
and monopole configurations considered on the same footing. The
results are: 1) though the correlation function of magnetic field
$b_{\mu}$ is affected by monopoles, the correlation function of
gauge potential $a_{\mu}$ is not, indicating that compact
QED$_{3}$ has the same perturbative (non-topological) structure
with non-compact QED$_{3}$; 2) CSB and confinement that is caused
by monopoles take place simultaneously in the half-filling
antiferromagnetic state; 3) both CSB and single monopoles are
suppressed in the $d$-wave superconducting state. We also argue
that the Anderson-Higgs (AH) mechanism can not confine matter
fields. The physical implications of these field theoretic results
are also discussed in the context.

We first consider the low-energy effective theory of Heisenberg
antiferromagnetism \cite{Ioffe, Marston89, Kim99}. The Lagrangian
is
\begin{equation}
{\cal L}_{F}=\sum_{\sigma=1}^{N}\overline{\psi}_{\sigma} \left(
\partial_{\mu }-ia_{\mu} \right) \gamma_{\mu}\psi_{\sigma} + \frac{1}{4}F_{\mu \nu}^{2}.
\end{equation}
The fermion $\psi_{\sigma}$ is a $4\times 1$ spinor and the $4
\times 4$ $\gamma _{\mu}$ matrices obey the algebra $\lbrace
\gamma_{\mu},\gamma_{\nu} \rbrace=2\delta_{\mu \nu}$. The Maxwell
term $F_{\mu \nu}=\partial_{\mu}a_{\nu}-\partial_{\nu}a_{\mu}$ is
kept here. It is convenient to write the action for the compact
gauge field $a_{\mu}$ as
\begin{equation}
S_{a} \propto \sum_{x,\mu,\nu}\left(F_{\mu \nu}-2\pi n_{\mu
\nu}\right) \left( \frac{1}{2}+\pi (q)\right) \left( F_{\mu
\nu}-2\pi n_{\mu\nu} \right),
\end{equation}
where the $n_{\mu \nu}$ are integers. The term $1/2$ comes from
the Maxwell term of $a_{\mu}$ and $\pi (q)$ is the vacuum
polarization of fermions. If the fermions are massless, the vacuum
polarization is $\pi (q)=N/8\left| q \right|$ to the one-loop
approximation \cite{Kim99}. At large distances this term dominates
and the term proportional to $1/2$ can be neglected. Then the
action for a gas of monopoles of charge $q_{a}=\pm 1$ is
\begin{equation}
S_{mono}=\frac{\pi^{2}N}{4}\sum_{a,b}q_{a}q_{b}V({\bf x_{a}}-{\bf
x_{b}}).
\end{equation}
The potential $V({\bf x})$ between monopoles is
\begin{equation}
V({\bf x})=\int \frac{d^{3}k}{(2\pi)^{3}}
\frac{e^{ikx}}{k^{3}}\sim \ln |{\bf x}|.
\end{equation}
Since the monopoles interact with a logarithmic potential, a
Kosterlitz-Thouless (KT) transition would take place at some
critical flavor $N_{cf}$ below which single monopoles are
proliferated. However, there is a controversy on $N_{cf}$. Some
authors found that $N_{cf}=24$ \cite{Ioffe, Kleinert} while others
found that $N_{cf}=0.9$ \cite{Marston}. We should emphasize that,
even if single monopoles are important, we can not immediately
draw the conclusion that massless fermions are confined. The
reason is that Wilson's area law \cite{Wilson} was proposed to
describe confinement of pure gauge field and static charge
sources. It loses its meaning when the gauge field couples to
dynamical massless fermions. When the sources separate, it becomes
more favorable to create a pair of fermions which then screens the
gauge force \cite{Fradkin}. However, if the massless fermions
acquire a finite mass via the CSB mechanism they then can be
considered as static sources. Indeed, when the fermions become
massive, creating a fermion-antifermion pair out of the vacuum
would cost a large amount of energy.

CSB can be studied by the standard Dyson-Schwinger (DS) equation
approach. The propagator of massless fermions is
$S^{-1}(p)=i\gamma \cdot p$. Interaction with gauge field
renormalizes it to $S^{-1}(p)=i\gamma \cdot p A \left( p^{2}
\right)+\Sigma \left( p^{2} \right)$ with $A(p^{2})$ the wave
function renormalization and $\Sigma (p^{2})$ the fermion
self-energy. The self-energy $\Sigma(p^{2})$ satisfies a set of
self consistent DS equations, which to the lowest order in $1/N$
expansion has the simple form
\begin{equation}
\Sigma(p^{2})=\int\frac{d^{3}k}{(2\pi)^{3}}\frac{\gamma^{\mu}D_{\mu\nu}(p-k)\Sigma(k^{2})\gamma^{\nu}}{k^{2}
+\Sigma^{2}(k^{2})},
\end{equation}
where a bare vertex is adopted \cite{Appelquist88}. If this
equation has only vanishing solutions, the gauge field is an
irrelevant perturbation and fermions remain massless. If $\Sigma
(p^{2})$ develops a nontrivial solution, the massless fermions
acquire a finite mass which breaks the chiral symmetry of
Lagrangian (1). For non-compact gauge field, the propagator in the
Landau gauge is
\begin{equation}
D_{\mu\nu}(q)=\frac{1}{q^{2}}\left(\delta_{\mu\nu}-\frac{q_{\mu}q_{\nu}}{q^{2}}\right).
\end{equation}
It was found \cite{Appelquist88} that CSB can only take place for
$N < N_{c} = 32/\pi^{2}$. The CSB in noncompact QED$_{3}$ has been
used to understand many properties of cuprate superconductors
\cite{Kim99, Tesanovic}.

To investigate the DS equation in compact QED$_{3}$, we should
first know the effect of monopoles on the propagator of $a_{\mu}$.
According to the arguments of Polyakov, the magnetic field
correlation function \cite{Polyakov77} is
\begin{equation}
\langle b_{\mu}b_{\nu} \rangle = \langle b_{\mu}b_{\nu}
\rangle_{0} + \langle b_{\mu}b_{\nu} \rangle _{m},
\end{equation}
where $\langle b_{\mu}b_{\nu} \rangle_{0}$ is the propagator
without monopoles and $\langle b_{\mu}b_{\nu} \rangle_{m}$ is the
contribution of monopoles. The density operator of monopoles is
$\rho (x)=\sum_{a} q_{a}\delta(x-x_{a})$ which is related to the
magnetic field as $b_{\mu}(x)=\frac{1}{2}\int d^{3}y
\frac{(x-y)_{\mu}}{\left|x-y\right|^{3}}\rho(y)$ or
$b_{\mu}(q)=\frac{q_{\mu}}{q^{2}}\rho(q)$ in the momentum space.
The singular contribution to the magnetic filed correlation
function is
\begin{eqnarray}
\langle b_{\mu}b_{\nu} \rangle _{m}=\frac{q_{\mu}q_{\nu}}{q^{4}}
\langle \rho(q) \rho(-q)
\rangle=\frac{q_{\mu}q_{\nu}}{q^{4}}\frac{M^{2}q^{2}}{q^{2}+M^{2}}.
\end{eqnarray}
Then we have
\begin{eqnarray}
\langle b_{\mu}b_{\nu} \rangle = \delta_{\mu \nu} -
\frac{q_{\mu}q_{\nu}}{q^{2}} +
\frac{q_{\mu}q_{\nu}}{q^{2}}\frac{M^{2}}{q^{2}+M^{2}}= \delta_{\mu
\nu} - \frac{q_{\mu}q_{\nu}}{q^{2}+M^{2}}.
\end{eqnarray}
The appearance of a pole at $-M^{2}$ was interpreted by Polyakov
\cite{Polyakov77} as the evidence of a finite mass gap for compact
gauge field.

In quantum gauge field theories, it is the gauge potential
$a_{\mu}$ that couples directly to dynamical matter fields.
Therefore, to study CSB we should calculate the propagator of
$a_{\mu}$. The magnetic field $b_{\mu}$ is related to $a_{\mu}$ as
$b_{\mu}=\epsilon_{\mu \nu \lambda}q_{\nu}a_{\lambda}$. Then the
correlation function of $a_{\mu}$ is
\begin{equation}
D_{\mu \nu} (q) = \langle a_{\mu}a_{\nu} \rangle = \epsilon_{\mu
ij}\epsilon_{\nu kl}\frac{q_{i}q_{k}}{q^{4}} \langle b_{j}b_{l}
\rangle.
\end{equation}
Using the fact that
\begin{equation}
\epsilon_{\mu ij}\epsilon_{\nu kl}\frac{q_{i}q_{k}}{q^{4}} \langle
b_{j}b_{l} \rangle_{m} = \epsilon_{\mu ij}\epsilon_{\nu
kl}\frac{q_{i}q_{k}q_{j}q_{l}}{q^{6}} \frac{M^{2}}{q^{2}+M^{2}}=0.
\end{equation}
Then we get the propagator
\begin{eqnarray}
D_{\mu \nu}(q) =\epsilon_{\mu ij}\epsilon_{\nu
kl}\frac{q_{i}q_{k}}{q^{4}} \langle b_{j}b_{l} \rangle_{0}
=\frac{1}{q^{2}}\left(\delta_{\mu \nu} -
\frac{q_{\mu}q_{\nu}}{q^{2}}\right).
\end{eqnarray}
It is clear that the monopole configurations do not affect the
propagator of $a_{\mu}$. Although magnetic field $b_{\mu}$ is the
quantity that can be detected directly by experiments, $a_{\mu}$
is the physical quantity that couples directly to matter fields.
Therefore, the monopole configurations do not affect the
interaction of $a_{\mu}$ with matter fields, at least within the
perturbation theory. This might not be too surprising if we note
the fact that $a_{\mu}$ always interacts $locally$ with matter
fields but the monopoles reflect the nontrivial topology of the
gauge field configuration which is certainly a $global$ property.

The results about CSB obtained in non-compact QED$_{3}$
\cite{Appelquist04} also applies to compact QED$_{3}$. If the
flavor of massless fermions $N < N_{c}$, CSB takes place, while if
$N > N_{c}$ the fermions remain massless and the chiral symmetry
is respected. Although there is a little debate on the value of
$N_{c}$, most analytic and numerical calculations indicated that
it is about $3.3 \sim 4$. For $N < N_{c}$, the massless fermions
becomes massive and hence its contribution to the vacuum
polarization is
\begin{equation}
\pi (q) =\frac{N}{4\pi} \left(\frac{2m}{q^{2}} +
\frac{q^{2}-4m^{2}}{q^{2}|q|} \arcsin
\left(\frac{q^{2}}{q^{2}+4m^{2}} \right)^{1/2} \right).
\end{equation}
Here we adopt a constant fermion mass $m$ for simplicity. We only
care about the behavior at very low momentum limit since
confinement is essentially a phenomenon of large distances. It is
easy to see that $\pi (q) \rightarrow N/8\pi m$ in the $q
\rightarrow 0$ limit. Obviously, the only effect of massive
fermions on the action of monopoles is a renormalization of the
gauge coupling constant. Consequently, the monopoles are in the
Coulomb gas phase, just like in the pure compact gauge theory.
Since fermions are massive there are undoubtedly no fermion zero
modes which, if exist, would suppress the monopole configurations
\cite{Marston}. The massive fermions can be approximately
considered as static charges. Then Wilson's confinement criteria
can be used safely. Confinement was found \cite{Polyakov77}
unambiguously after calculating the Wilson loop $F[C] = <
e^{i\oint a_{\mu}dx_{\mu}} >$. Therefore, CSB and confinement take
place simultaneously for $N < N_{c}$ \cite{note}. For $N
> N_{c}$, a careful analysis of KT transition and deeper insights
on the criteria of confinement for massless fermions are needed
\cite{Hermele}.

For cuprate superconductors, the physical flavor is $N=2 < N_{c}$,
corresponding to the two components of spin $1/2$. Then CSB and
confinement both occur at half-filling and prevent the appearance
of mobile fermions at low temperatures. This is consistent with
the fact that the low temperature thermal conductivity vanishes at
very low doping concentrations \cite{Sutherland}. CSB generates a
finite gap for the gapless nodal fermions. This accounts for the
finite nodal gap observed by angle-resolved photoemission
spectroscopy (ARPES) measurements in lightly doped cuprates
\cite{Shen}. Moreover, when CSB happens, a long-range
antiferromagnetic order is formed, corresponding to the well-known
N\'{e}el order of undoped cuprates \cite{Kim99, Tesanovic}.

We next would like to consider confinement of matter fields in the
$d$-wave superconducting state at finite doping concentration. To
describe the $d$-wave superconductor, we should couple both
massless Dirac fermions and holons $\phi$ to the U(1) gauge field
$a_{\mu}$. The scalar field $\phi$ develops a nonzero vacuum
expectation ($\left <\phi\right > \neq 0$) and the gauge boson
$a_{\mu}$ acquires a finite mass $\xi$ via AH mechanism. The gauge
boson mass $\xi$ suppresses CSB completely \cite{Liu2003}, hence
the low-energy excitations are gapless nodal fermions. (Note that
there should not be a Yukawa coupling term $\phi \overline {\psi}
\psi$ between massless fermions and scalar field $\phi$. If such a
term were present, then the nonzero $\left <\phi\right >$ would
generate a finite mass for the massless fermions, in disagreement
with experiments).

Using a simple but compelling argument, we can show that single
monopoles can not exist in a superconductor. When a single
monopole is placed in the interior of a superconductor its line of
magnetic flux must have somewhere to go. However, according to the
Meissner effect, a superconductor always repels the magnetic
field. Thus the magnetic flux emitting from a monopole must end at
an anti-monopole. In other words, the monopoles must appear in the
form of bound pairs and all the magnetic flux is trapped into a
thin tube. To see this more explicitly, we can calculate the
potential between two monopoles in a superconductor. The
propagator of massive gauge boson is
\begin{equation} D_{\mu\nu}(q) = \frac{1}{q^{2}
\left[1+\pi(q^{2})+\xi^{2}q^{-2}
\right]}\left(\delta_{\mu\nu}-\frac{q_{\mu}q_{\nu}}{q^{2}}\right).
\end{equation}
The gauge boson mass term $\xi^{2}q^{-2}$ dominates the low
momentum behavior, no matter whether the fermions are massless or
not. Using the same calculations that lead to (4), we found a
linear potential
\begin{equation}
V({\bf x})=\int \frac{d^{3}k}{(2\pi)^{3}}
\frac{e^{ikx}}{k^{4}}\sim |{\bf x}|
\end{equation}
between monopoles. Therefore, single monopoles can not exist and
there is a string between a monopole and an anti-monopole. The
superconductor can be understood by the picture that condensation
of charged particles gives rise to confinement of magnetic
monopoles. If we interchange the roles of electricity and
magnetism, then we get a dual picture that condensation of
magnetic monopoles causes confinement of charged particles, which
describes the half-filled antiferromagnetic state. Thus an
"electromagnetic" duality exist between the Heisenberg
antiferromagnet and the superconductor, which might help us to
understand the physics of cuprate superconductors. This kind of
duality also underlies the most exciting attempts \cite{Hooft}
made recently towards a final understanding of quark confinement
in QCD.

The spin-charge separation and recombination have been studied
extensively \cite{Kim99, Wen96, Mudry, Rantner}. It is generally
expected that spinons and holons are bound together to form real
electrons in the $d$-wave superconductor. Two possible ways have
been proposed \cite{Rantner} to realize the confinement: AH
mechanism and monopole effect. Since single monopoles do not exist
in a superconductor, it seems natural that it is the AH mechanism
that causes confinement. However, we believe that this is not the
case. Remember that Higgs mechanism (the non-Abelian
generalization of Abelian AH mechanism) appears in the standard
model of electro-weak interaction \cite{Weinberg}. Although the
intermediate gauge bosons acquire finite mass gap, the fermions
and the gauge bosons are certainly not confined. Confinement via
AH mechanism requires that the gauge coupling must be very strong
at the $q \rightarrow 0$ limit. For QED$_{3}$, we can define a
dimensionless running gauge coupling $\overline{\alpha} (q)$ as
\cite{Appelquist04}
\begin{equation} \overline{\alpha}(q) =
\frac{e^{2}q}{q^{2}+\xi^{2}+(e^{2}N/8)q} = \frac{8}{N}\frac{\alpha
q}{q^{2}+\xi^{2}+\alpha q}.
\end{equation}
The running coupling constant $\overline{\alpha} (q)$ vanishes at
both $q \rightarrow \infty$ and $q \rightarrow 0$ limits. Since
the gauge coupling is weakened by the gauge boson mass generated
via AH mechanism, the matter fields should not be confined. We can
make a comparison between the coupling strengths that are needed
to cause CSB and confinement. Suppose that CSB takes place, then
the potential between a fermion and an anti-fermion has a
logarithmic form \cite{Maris}, $V({\bf x})\sim \frac{\ln\left|{\bf
x}\right|}{1+\pi(0)} \sim \ln\left|{\bf x}\right|$, with $\pi(0)$
the vacuum polarization of massive fermions at zero momentum. But
in general confinement requires a linear potential $V({\bf x})\sim
\left| {\bf x}\right|$ between two particles. The attractive force
that is needed to cause confinement should be much stronger than
that needed to cause CSB. In general, when the gauge boson
acquires a finite mass via AH mechanism, its coupling is not
strong enough to cause CSB \cite{Liu2003}. Thus it certainly can
not cause confinement.

We now see that both monopole effect and AH mechanism can not be
the confining mechanism that leads to spin-charge recombination.
This leaves us with two possibilities: confinement is caused by a
new unknown mechanism or it does not occur in the superconducting
state. The later possibility is not impossible. At present, almost
all ARPES experiments supporting the existence of well-defined
quasiparticle peaks in superconducting state have been performed
in the $(\pm \pi, 0)$ directions \cite{Orenstein}. Recent ARPES
experiments in the $(\pm \pi/2, \pm \pi/2)$ directions revealed a
much shorter quasiparticle lifetime than that was predicted by
BCS-like theory \cite{Orenstein, Valla}. Other evidence for the
existence of well-defined quasiparticles comes from the finite
thermal conductivity at low temperatures observed by heat
transport measurements \cite{Taillefer}. However, the heat
transport behavior could also be described by the spinons and
spin-charge recombination is not required. To tell which
possibility actually works needs further investigations.

It was showed \cite{Fradkin} that the Higgs and confining phases
are smoothly connected when the Higgs fields transform like the
fundamental representation of the gauge group. While this result
applies to other lattice gauge theories, it does not apply to the
compact QED$_{3}$. In the Higgs phase, a true gauge boson mass gap
is generated by vacuum degeneracy, while in the confining phase
there is no vacuum degeneracy and the monopoles only affect the
correlation function of $b_{\mu}$, leaving that of $a_{\mu}$
unchanged.

The results in this paper can understand some physics of
high-$T_{c}$ superconductors from a field theoretic point of view
and are also helpful in studying confinement of more complicated
gauge theories such as QCD.

This work is supported by the NSF of China No. 10404024 and the
China Postdoctoral Science Foundation.


\begin{thebibliography}{99}
\bibitem{note}
The kinetic term $F^{2}_{\mu \nu}$ should be taken away when we
consider the effective theory of cuprate superconductors
\cite{Ioffe, Marston89, Kim99}. However, it is easy to see that
the conclusion does not change.
\end{thebibliography}
\end{document}